\documentclass[1p,final]{elsarticle}
\usepackage{amsmath}
\usepackage{amssymb}
\usepackage{amsfonts}
\usepackage{dsfont}
\usepackage{MnSymbol}
\usepackage{lineno}  
\usepackage[english]{babel}
\usepackage{slashed}
\usepackage{bm,braket}
\usepackage{verbatim}
\usepackage{booktabs}
\usepackage{mathtools}
\usepackage{graphicx}
\usepackage[pdftex,dvipsnames,table]{xcolor}
\usepackage{graphbox}
\usepackage{bigstrut}
\usepackage[T1]{fontenc}
\usepackage{nicefrac}
\usepackage{float}
\usepackage{braket}
\usepackage[export]{adjustbox} 
\usepackage{makecell}
\usepackage[normalem]{ulem}
\usepackage[section]{placeins}
\usepackage{tabularx}
\usepackage{xargs}      
\journal{Progress in Particle and Nuclear Physics}
\biboptions{sort&compress}
\usepackage{titlesec}
\usepackage{sectsty}
\titleformat{\section}{\normalfont\Large\bfseries}{\thesection}{1em}{}
\titleformat{\subsection}{\normalfont\large\bfseries}{\thesubsection}{1em}{}
\titleformat{\subsubsection}{\normalfont\normalsize\bfseries}{\thesubsubsection}{1em}{}
\usepackage{hyperref}
\hypersetup{,  
	colorlinks= true,
	urlcolor  = blue,
	linkcolor = blue,
    urlcolor  = blue,
    citecolor = blue,
    anchorcolor = blue,
	pdftitle  = {Review on dynamical coupled-channel approaches},
	pdfauthor = {M. Doering, J. Haidenbauer, M. Mai, T. Sato},
	pdfsubject= {Review on dynamical coupled-channel approaches}
}
\usepackage{cleveref}
\crefformat{figure}{Fig.~#2#1#3}
\crefformat{table}{Tab.~#2#1#3}
\crefformat{equation}{Eq.~(#2#1#3)}
\crefformat{section}{Sect.~#2#1#3}
\numberwithin{equation}{section}
\graphicspath{{plots/}} 

\newcommand\vertarrowbox[3][8ex]{%
  \begin{array}[t]{@{}c@{}} #2 \\
  \left\updownarrow\vcenter{\hrule height #1}\right.\kern-\nulldelimiterspace\\
  \makebox[0pt]{\scriptsize#3}
  \end{array}%
}

\newcommand{\GeV}{{\rm GeV}}
\newcommand{\MeV}{{\rm MeV}}

\renewcommand{\Re}{{\rm Re}\,}

\newcommand{\dv}[1]{\mathrm{d} #1}
\newcommand*\dif{\mathop{}\!\mathrm{d}}

\newcommand{\be}{\begin{eqnarray}}
\newcommand{\ee}{\end{eqnarray}}
\newcommand\TT{\rule{0pt}{2.6ex}}       
\newcommand\BBB{\rule[-1.6ex]{0pt}{0pt}} 

\newlength{\feynwidthb} 
\newcommand{\ba}{\begin{eqnarray}}
\newcommand{\ea}{\end{eqnarray}}

\makeatletter
\def\widebreve{\mathpalette\wide@breve}
\def\wide@breve#1#2{\sbox\z@{$#1#2$}%
     \mathop{\vbox{\m@th\ialign{##\crcr
\kern0.08em\brevefill#1{0.8\wd\z@}\crcr\noalign{\nointerlineskip}%
                    $\hss#1#2\hss$\crcr}}}\limits}
\def\brevefill#1#2{$\m@th\sbox\tw@{$#1($}%
  \hss\resizebox{#2}{\wd\tw@}{\rotatebox[origin=c]{90}{\upshape(}}\hss$}
\makeatletter

\begin{document}
\begin{frontmatter}
\title{Phenomenology of light baryon resonances}

\author[1,2]{Michael D\"oring}
\ead{doring@gwu.edu}

\affiliation[1]{
    organization={Department of Physics, The George Washington University},
    addressline={725 21st St, NW}, 
    city={Washington},
    postcode={20052}, 
    state={District of Columbia},
    country={USA}}
\affiliation[2]{
Theory Center, Thomas Jefferson National Accelerator Facility, Newport News, VA 23606, USA
}

\begin{abstract}
Protons and neutrons are the  building blocks of matter, glued together in nuclei by strong interactions. They can be excited by pions, real and virtual photons, neutrinos and other probes. These excitations are referred to as light baryon resonances. A short, pedagogical overview of the field is presented including experimental progress, interpretation of light baryon resonances, and a focus on analysis methods to extract the resonance spectrum from data.
\\~\\
Preprint number: JLAB-THY-25-4577.
\end{abstract}

\begin{keyword}
Light baryon resonances \sep 
Baryon spectroscopy \sep 
Amplitude analysis \sep
S-matrix theory \sep
Photoproduction reactions\sep
Dynamical coupled-channel models

\end{keyword}
\end{frontmatter}

\section{Background}
Light baryon resonances are emergent phenomena in particle scattering involving one nucleon. They represent excited states of protons and neutrons. They are considerable heavier than the ground-state nucleons but their energy is still significantly below the high-energy regime. At high energies, quarks and gluons are the particles that interact strongly, as manifested in the production of two- and three-prong jets in high-energy collisions representing one of most direct evidence for these particles. Strong interactions still govern the formation of nuclei at very low energies, but the dynamics is fundamentally different. Here, quarks are confined in protons and neutrons, and the interaction of these fundamental building blocks of matter is mediated by the exchange of pions that are themselves composite objects. The same underlying theory of strong interactions, Quantum Chromodynamics (QCD), governs both regimes. Not only protons and neutrons are emergent phenomena of QCD, but also light baryon resonances. Situated in the transition region between high and low energies they have attracted the continued interest of nuclear physics for being a key to our understanding of matter formation. 

Quark models, Dyson-Schwinger approaches, holographic models, and other formulations predict the resonance spectrum with the intention of a direct connection to QCD in terms of quarks and gluons. Other approaches rely on the mesons and baryons as building blocks whose interactions can generate some resonances. Lattice QCD begins to predict the light baryon spectrum from QCD directly. On the experimental side, the last two decades have seen an unprecedented activity in light baryon spectroscopy motivated by these predictions but also by questions related to $S$(scattering)-matrix theory such as complete experiment, and hadron dynamics leading to non-resonant phenomena such as cusps and triangle singularities. 

In this article, we mention both the theory and experimental aspects of light baryons but focus on the phenomenological aspects of how the light baryon spectrum can be extracted from data. This work summarizes parts of a recent review~\cite{Doring:2025sgb}. A review on the interpretation of the light baryon resonance spectrum has appeared recently~\cite{Burkert:2025coj}. Another recent review on the experimental aspects is provided in Ref.~\cite{Thiel:2022xtb}. A theory review of resonances in general can in be found in Ref.~\cite{Mai:2022eur}. Reference~\cite{Doring:2025sgb} provides a literature collection of almost all reviews and other key publications on light baryon spectroscopy.
\section{Resonances -- a quantum mechanical example}
\label{sec:qm}
To access the topic at the advanced undergraduate level, we illustrate how resonances appear in potential scattering. 
A follow-up example of a resonance in a simple $\phi^4$ quantum field theory is given in Ref.~\cite{Mai:2025wjb}. Another pedagogical follow-up read can be found in Ref.~\cite{Habashi:2020ofb} in which an effective field theory with a resonance field are used to illustrate resonances and their renormalization.

We consider the finite spherical well with variable depth $v_0$
\be
V(r)=\begin{cases}
	-v_0&r<R,\\
	0 &r\ge R \ .
\end{cases}
\ee
The solution $\cot \delta_\ell$ of the scattering equation for $\ell$-wave is known analytically~\cite{Sakurai_Napolitano_2020} and often determined in an advanced undergraduate course on quantum mechanics by logarithmically matching the wave function at the well wall.

One can also solve the
Lippmann-Schwinger equation (LSE) to get the phase shift. In momentum space and for $\ell$-wave it reads
 \begin{equation}
	T_{\ell}(p',p)=V_{\ell}(p',p)+\int\limits_0^{\infty}\dif q\,q^2\,\frac{V_{\ell}(p',q)}{E-\frac{q^2}{2m}+i\epsilon}\,T_{\ell}(q,p) \ .
	\label{eq:7.179}
\end{equation}   
The partial-wave projected $s$-wave potential reads
\begin{equation}
	V_\ell(p',p)=\frac{2}{\pi \hbar^3} \int\limits_{0}^{\infty}
	\dif r\, r^2\,j_\ell(k'r)V(r)\,j_\ell(kr) \ ,
	\label{eq:pwa}
\end{equation}
where $j_\ell$ are the spherical Bessel functions, $p=\hbar k$, and
\be
V_0(p',p)
=
\frac{2}{\pi\hbar^3}\,\frac{-v_0}{k'k}\,f(k',k)\quad\text{with}\quad f(k',k)=
\frac{k'\sin(kR)\cos(k'R)-k\cos(kR)\sin(k'R)}{k^2-k'^2} \ ,
\label{eq:V0}
\ee
with well-defined limits $k'\to k$ and $k\to 0$. The connection of $T_\ell$ to scattering amplitude and phase shifts is given by
\be
 f(\theta)=\sum_{\ell=0}^\infty (2\ell+1)\,t_\ell\, P_\ell(\xi) \ , \quad
t_\ell=\frac{1}{k\cot\delta_\ell-ik}=-\pi m\hbar T_\ell (p_\text{cm},p_\text{cm}) \ ,
\label{eq:tanddelta}
\ee
with Legendre polynomials $P_\ell$, $\xi=\cos\theta$ the scattering angle, and on-shell momentum $p_\text{cm}=\sqrt{2mE}$.
 There are different numerical methods to solve the LSE. One is to discretize the momentum integration into $q_i,\,i=1,\dots,n$ with weights $w_i$ leading to the matrix equation
\begin{equation}
		\bar{T}=\bar{V}+\bar{V}\bar G\bar T \,\,\Rightarrow\,\,
	\bar{T}=(\mathds{1}-\bar{V} \bar{G})^{-1}\bar{V} \ ,
	\label{eq:disclse}
\end{equation}
 where index $\ell$ and argument $E$ are suppressed.
 For $\bar A$ representing either $\bar V$ or $\bar T$, $\bar A\in\{\bar V, \bar T\}$ one can show that 
\ba
\bar A&=&
\begin{pmatrix}
	A_{11}		& \cdots	& A_{1n}	& A_{1,n+1}	\\
	\vdots		& \ddots	& \vdots	& \vdots	\\
	A_{n1}		& \cdots	& A_{nn}	& \vdots	\\
	A_{n+1,1}	& \cdots	& \cdots	& A_{n+1,n+1}
\end{pmatrix}, \,\,
\bar G=
\begin{pmatrix}
	\frac{q_1^2\,w_1}{E-\frac{q_1^2}{2m}}	&0	&\cdots	&0	\\
	0				&\ddots &	&\vdots	\\
	\vdots				&	&\frac{q_n^2\,w_n}{E-\frac{q_n^2}{2m}}&\vdots \\
	0				& \cdots&\cdots&0
\end{pmatrix}, 
\label{eq:ingredients}
\ea
leads indeed to a discretized version of the LSE, with on-shell momentum $q_{n+1}=q_\text{cm}(E)$ that allows to determine the on-shell transition $T (p_\text{cm},p_\text{cm})$. To avoid the two-body singularity at $q_\text{cm}$ in \cref{eq:7.179} one can deform the integration contour:
\begin{center}
	\includegraphics[width=0.4\textwidth]{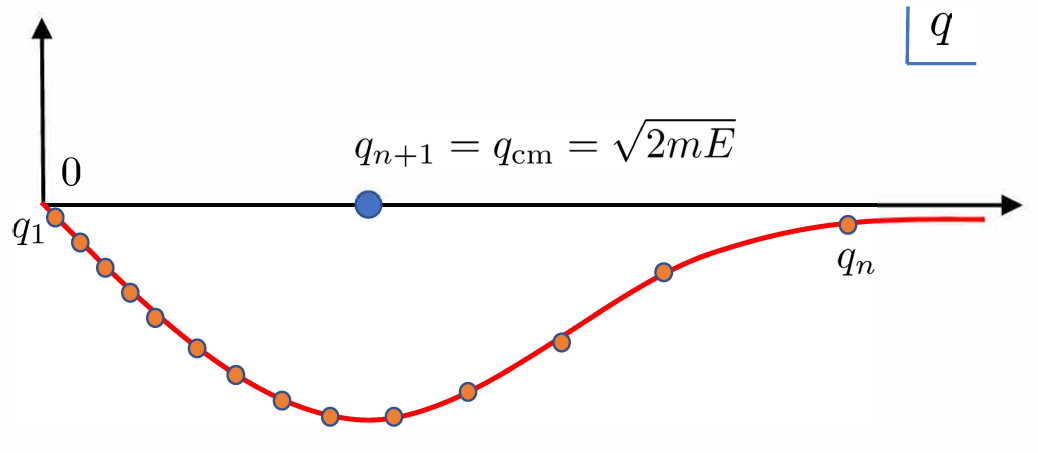}	
\end{center}
For this particular potential with $f$ in Eq.~\eqref{eq:V0}, the potential has to bend back to the real axis to ensure convergence while in general other paths are possible, see the discussion in Ref.~\cite{Doring:2025sgb}.

\begin{figure}[tb]
    \centering
    \includegraphics[width=0.43\textwidth]{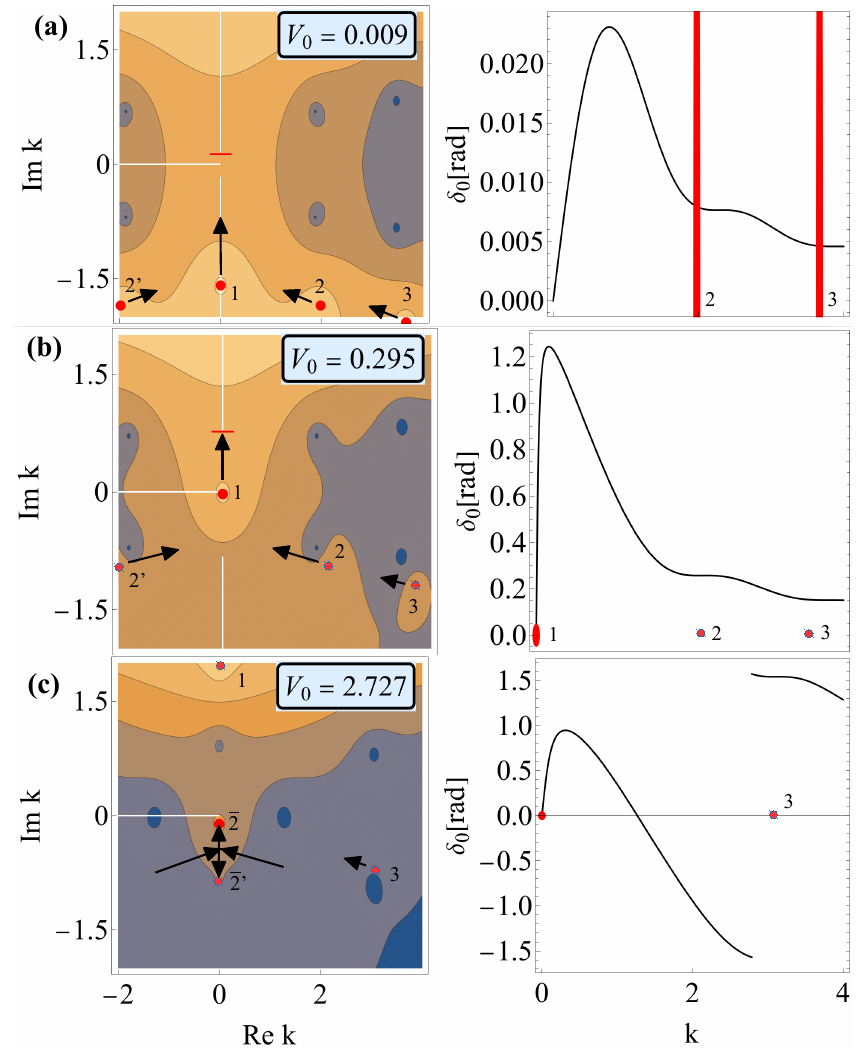}
    \includegraphics[width=0.26\textwidth]{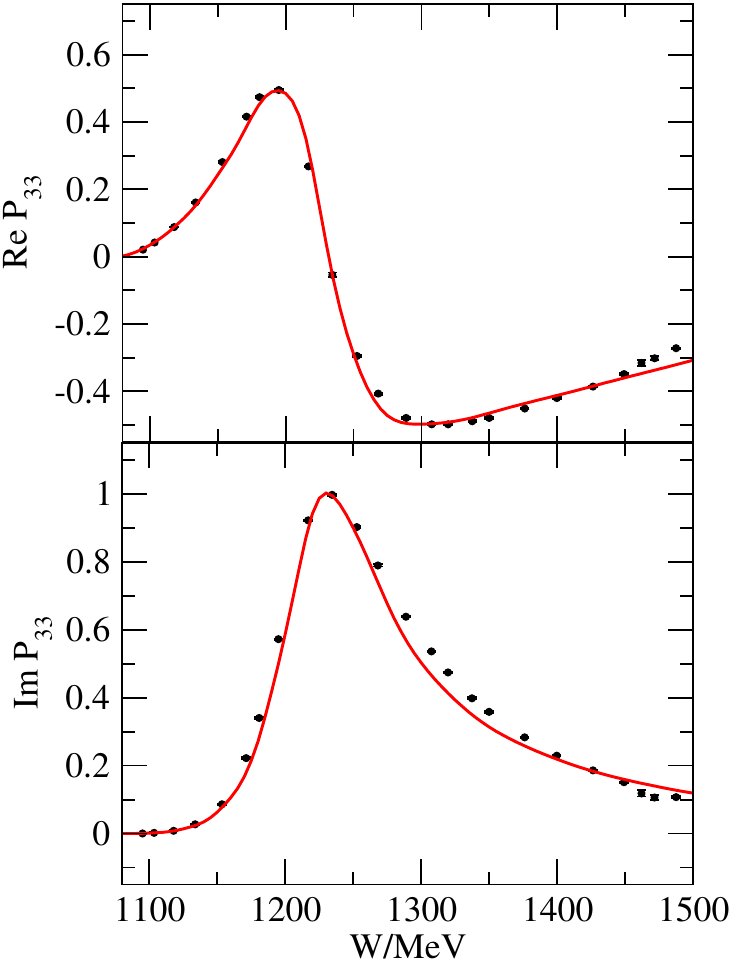}
    \includegraphics[width=0.26\textwidth]{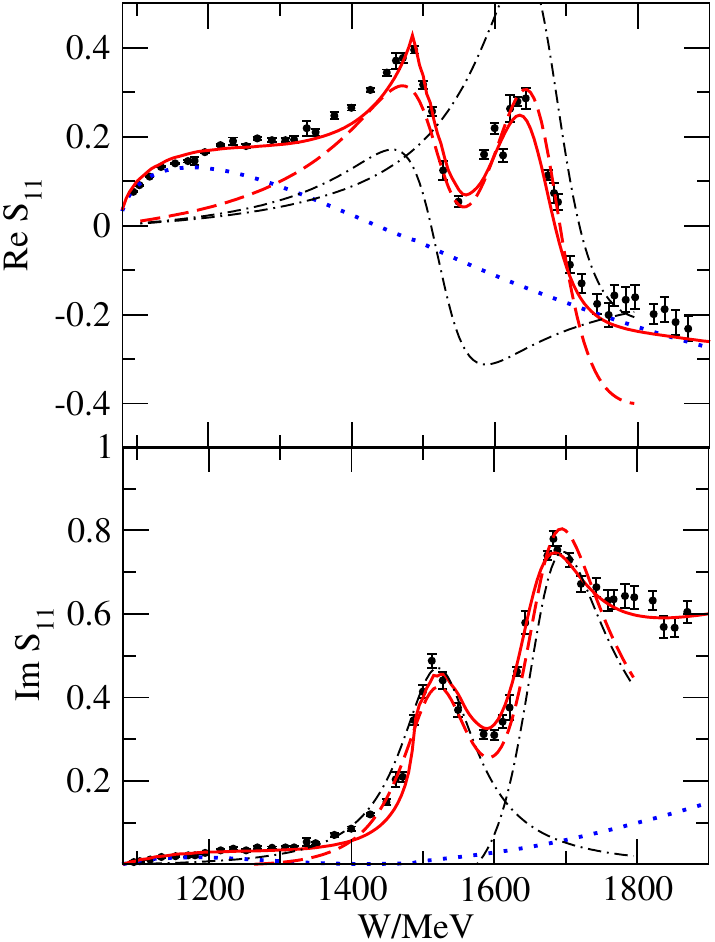}
    \caption{{\bf Left:} Scattering and bound states of the finite spherical well.  The screenshots a)-c) show the situation for increasing potential depth, see also an online animation~\cite{Doering:blog}.
The left column shows the $S$-wave $T$-matrix, $|t_0|$, in the complex-momentum $k$-plane (arb. units), 
the right column shows the phase shift. (a) For a shallow potential, there is no bound state, but only virtual state 1 and resonances 2 and 3. In (b), the scattering length is much larger than the dimension of the potential (universality). In (c), pole 1 became a deeply bound state. Pole 2 and its mirror pole 2' have met on the imaginary $k$-axis and then separated again as virtual states $\bar 2$ and $\bar 2'$, 
with $\bar 2$ on its way to become a bound state and $\bar 2'$ a deeper-bound virtual state.
{\bf Center:} The $P_{33}$ partial wave and the $\Delta(1232)3/2^+$ resonance. The data for this and the right picture (taken from Refs.~\cite{Doring:2009bi, Doring:2009yv}) represent the Single-Energy-Solutions (SES) from the SAID data base~\cite{SAID-web}. 
{\bf Right:} The $S_{11}$ partial wave. In contrast to the $P_{33}$ partial wave, there are the two overlapping resonances $N(1535)1/2^-$ and $N(1650)1/2^-$ on top of a substantial background (compare to the spherical well). In addition, there is a sharp threshold cusp at $W=1486$~MeV next to the $N(1535)1/2^-$. See text for further explanations.
    }
    \label{fig:lamandf0}
\end{figure}

The $S$-wave phase shift from Eq.~\eqref{eq:tanddelta} is shown in the second column of Fig.~\ref{fig:lamandf0} for different potential depths.
An animation of the process as $V_0=v_0$ increases can be found online~\cite{Doering:blog}.
The figure caption explains the transition of resonance poles to bound state. Resonance poles like pole 2 come with a mirror pole, $2'$. In the variable $E=k^2/(2m)$ the amplitude has two Riemann sheets. The poles closer to the physical axis ($k>0$) are visible as small structures in the phase shift. The mirror pole is also located on the second Riemann sheet but at Im~$E>0$ (or third quadrant in the $k$-plane). The fusion of both poles is shown in the lower left panel - after the poles meet, one stays on the second Riemann sheet but below threshold (pole $\bar 2'$). This is referred to as ``virtual state''. The other pole, $\bar 2$ moves to the threshold at $k=0$, causing a  scattering length much larger than the dimension of the well. In this situation, referred to as universality~\cite{Braaten:2004rn}, the physical properties of the system become independent of the microscopic details of the potential. When the potential becomes even deeper, the pole $\bar 2$ finally becomes a bound state, moving along the positive Im~$k$ axis. 
This intriguing behavior of $S$-wave poles is the same as found for the pion-mass dependent pole trajectory of the $\sigma$ meson~\cite{Hanhart:2008mx}. 

To conclude, the discussed example even allows to illustrate Levinson's theorem that relates the phase shift in the limit $k\to\infty$ and the number of bound states:
$    \delta_\ell(0)-\delta_\ell(\infty)=n_\ell\,\pi \ .
$
For example, in panel c), $\delta(\infty)=-\pi$ and there is, indeed, one bound state (pole 1).

This quantum scattering example illustrates how bound states and scattering states are connected through analyticity of the scattering amplitude. It also demonstrates common problems regarding the inverse problem of extracting resonance information from data, i.e., the determination of light baryon resonances from measurements. 

For this, consider first the center and right-hand panels in Fig.~\ref{fig:lamandf0}. The center panel shows the $P_{33}$ partial wave for elastic pion-nucleon scattering in the notation $L_{2I2J}$ with $J$ the total angular momentum/spin of the resonance, $L$ the orbital angular momentum of the $\pi N$ system, and isospin $I$. The most prominent light baryon resonance, the $\Delta(1232)3/2^+$, is clearly visible (the $3/2^+$ indicates $J^P$). It is a textbook example of a Breit-Wigner (BW) resonance. In particular, Im~$P_{33}$ peaks with a value of 1 at the resonance position, and Re~$P_{33}$ has a zero. Equivalently, $\delta(P_{33})=\pi/2$ for a fully elastic resonance.
For more information on Breit-Wigner resonance parametrizations and other commonly used tools of resonance analysis like Argand plots, see, e.g., review 50 of the 2021 edition of the Particle Data Book~\cite{ParticleDataGroup:2020ssz}. 
The $\Delta(1232)3/2^+$ is the only light baryon resonance that is clearly and directly visible in pion- and photon-induced reactions, by being very strong and well separated from other states of different $J^P$.

In contrast to the BW shape of the $\Delta$, the resonances in the finite square well only show as rather small modulations in the phase shift. We note that the resonance poles  2 and 3 can be farther away from the real-$k$ physical axis than their distance in Re~$k$, see red dots in the left column. One refers to them as ``overlapping'' resonances that are difficult to isolate. In light baryon spectroscopy there are several such cases as the right panel in Fig.~\ref{fig:lamandf0} shows: not only are the the $N(1535)1/2^-$ and $N(1650)1/2^-$ broad and overlapping, but there is the nearby threshold of the $\eta N$ channel that further complicates the determination of the $N(1535)1/2^-$ resonance parameters. The black dash-dotted lines show the contribution to the amplitude from the resonance poles in terms of the residue, i.e., $a_{-1}/(W-W_0)$ where $W_0$ is a complex resonance pole position (in contrast to quantum mechanics, energy in light baryon spectroscopy is often denoted as $W$). Obviously, one resonance  provides a structured background to the other resonance. The sum of both contributions is shown with the red dashed lines. There are also contributions from non-resonant hadronic processes from exchanges in the $t$ and $u$-channel, referred to as non-pole $T$-matrix (dotted blue lines). This separation is, however, model-dependent as shown in Ref.~\cite{Doring:2009bi}.
Note also the recent discussion on an $S_{11}$ virtual state at very low energies and far in the complex energy plane, referred to as $N(920)$~\cite{Wang:2017agd, Li:2025man, Hoferichter:2015hva, Cao:2022zhn, Doring:2009uc}. In the above picture, an enhancement in Re~$S_{11}$ at threshold is visible.

The quantum scattering example and its comparison with $S_{11}$ and $P_{33}$ point out common problems in the phenomenology of light baryons (apart from the $\Delta(1232)3/2^+$ case): 
\begin{enumerate}
\item
The light baryon spectrum contains overlapping resonances with the same $J^P$. \item 
It also has resonances at similar mass but with different $J^P$ or isospin. For the above example, the region of $W\approx 1.5-1.6$~GeV is referred to as ``second resonance region'' owing not only to the presence of $J^P=1/2^-$ resonances but also to states like the $N(1520)3/2^-$. A separation of these states is only possible after the decomposition of the amplitude into partial waves. This corresponds to Eq.~\eqref{eq:tanddelta} but using Wigner-d functions instead of Legendre polynomials, allowing for the treatment of half-integer spin, see Sec.~2.5 of Ref.~\cite{Doring:2025sgb}. 
\item
The separation of the amplitude into partial waves requires experimental information beyond differential cross sections $d\sigma/d\Omega$. For $\pi N$ scattering, one must measure four quantities including different polarization observables~\cite{Bransden2015-vz} at the same energy and angle to determine the amplitude. For photo- and electroproduction reactions, the question of how many different observables are needed to determine not only the amplitude but also the partial-wave content is referred to as ``complete experiment'' and ``truncated partial-wave complete experiment'', respectively~\cite{Thiel:2022xtb}. As observables are bilinear (and not linear) combinations of amplitudes there are, in general, more measurements necessary as a naive counting of degrees of freedom suggests, due to discrete ambiguities~\cite{Workman:2016irf}.
\item 
Even if partial waves can be determined reliably, the example of the overlapping $J^P=1/2^-$ resonances demonstrates that isolating resonances from a background is difficult. Also, one finds the same resonance in different reactions, for example: $\pi N\to\eta N,\,K\Lambda,\,\pi\pi N$ but also the $\gamma N$ initial state combined with these different final states, referred to as (meson) photoproduction. In different reactions, the background changes. The only universal resonance parameters are complex pole positions $W_0$ and residues of different channel transitions $A\to B$, $a_{-1}^{A\to B}$. Here, Re~$W_0$ corresponds to the resonance mass, $-2\text{Im} W_0$ to the width, and the residues themselves to (transition) branching ratios~\cite{ParticleDataGroup:2024S}. The residues factorize, $a_{-1}^{A\to B}=g_Ag_B$ and the $g$ may be compared to traditional Breit-Wigner branching ratios after attaching suitable kinematic factors~\cite{Workman:2013rca}.
\end{enumerate}
To detect as many new resonances as possible, and to reliably confirm the known ones, recent years have seen an unprecedented surge in experimental and phenomenological activity to not only study pion-induced reactions, but also photo- and electroproduction production of one or more (vector) mesons. The motivation to measure different initial and final states is to become more sensitive to resonances that couple only weakly to the $\pi N$ state. Some of these activities will be discussed below.

\section{Theory approaches for resonances}
For a recent review of theory approaches for excited baryons see Ref.~\cite{Burkert:2025coj}. Their figure 2.1 contains also a compilation of phenomenologically extracted states from different groups. 
The minimal content for ground-state and excited baryons are three quarks. Quark models~\cite{Isgur:1977ef} usually impose a confinement potential similar to the mentioned spherical well, but in a generalization to three particles and variations regarding the residual spin-dependent interaction between the quarks. The confining potential is usually rising linearly with distance representing the string tension. 
Relativistic kinematic was incorporated early~\cite{Capstick:1986ter}. One puzzling finding of this model was that the Roper resonance $N(1440)1/2^+$~\cite{Roper:1964zza} turned out to be heavier than its parity partner, the $N(1535)1/2^-$. Chiral constituent quark models impose a confining potential and meson-like interactions derived from chiral symmetry~\cite{Glozman:1997ag}.
Quark models were further refined  in  relativistic  models with
instanton-induced quark forces   ~\cite{Loring:2001kx},  quark-diquark models~\cite{Santopinto:2004hw}, hypercentral constituent quark models~\cite{Giannini:2001kb}, in covariant models~\cite{Gross:2006fg, Ramalho:2008ra} with predictions of electromagnetic properties including timelike form factors~\cite{Ramalho:2012ng}. See also Refs.~\cite{ Vijande:2004he, Huang:2005gw, Garcilazo:2007eh, Bicudo:2016eeu, Zhong:2024mnt} for other quark models and related approaches for excited baryons. The relativistic light-front quark model and its application to electroproduction reactions was revied by Azauryan and Burkert~\cite{Aznauryan:2011qj} and by Aznauryan, Bashir, Braun, Brodsky, and Burkert in Ref.~\cite{Aznauryan:2012ba}. 
A graphical overview of different quark model predictions for light baryonic resonances in provided in Fig. 4.9 of Ref.~\cite{Burkert:2025coj}.
New perspectives of exploring hadron structure through transition GPDs were recently summarized in a whitepaper~\cite{Diehl:2024bmd}.

Holographic QCD is an approach with extra spatial dimensions to model hadrons, as explained in an introductory text by Erlich~\cite{Erlich:2014yha}. A suggestive example of a particle propagating in a compact extra dimension is provided to illustrate that excited  hadrons can be interpreted as Kaluza-Klein modes. 
Holographic QCD for baryon resonances has been proposed in Refs.~\cite{Erlich:2005qh, deTeramond:2005su}, see also Refs.~\cite{Karch:2006pv, deTeramond:2008ht}. The
approach of Ref.~\cite{deTeramond:2005su} is highly successful in organizing the hadron, see also the review of Ref.~\cite{Brodsky:2014yha}.

Three-quark dynamics is also studied in Dyson-Schwinger equations that are a coupled set of integral equations relating the different $n$-point functions of QCD to each other in a self-consistent way, including quark and gluon propagators and QCD vertices. See, e.g., Fig. 4.12 in Ref.~\cite{Burkert:2025coj}. Baryons appear as poles in the three-quark correlation functions. Coupling the baryons to the continuum to address resonance widths and incorporate hadron dynamics is in active development.
Few-quark dynamics was even extended to five-body systems~\cite{Eichmann:2025gyz}. For more recent reviews on the topic see Refs.~\cite{Ding:2022ows, Roberts:2021nhw, Burkert:2017djo, Eichmann:2016yit}.

The Roy-Steiner set of equations were applied to the $\pi N$ and other sectors in Ref.~\cite{Ditsche:2012fv, Hoferichter:2015dsa, Hoferichter:2015tha, Hoferichter:2015hva} and also used to extract resonance pole parameters including the $\Delta(1232)$ and Roper resonances~\cite{Hoferichter:2023mgy}. Furthermore, the authors confirm a sub-threshold singularity on the unphysical sheet in $S_{11}$ close to the circular cut, see \cref{sec:qm}.


\subsection{Lattice QCD and finite volume aspects}
\label{sec:lQCD}
Lattice QCD provides information about the spectrum, structure, and interactions of hadrons as they emerge from quark-gluon dynamics. This numerical approach relies on Monte-Carlo sampling of correlation functions on Euclidean space-time evaluated in a finite (space-time) volume. Due to this, real-time observables such as scattering amplitudes cannot be accessed directly. Instead, as pioneered by L\"uscher for two-body systems~\cite{Luscher:1985dn, Luscher:1986pf, Luscher:1990ux}, the dependence of energy eigenvalues on the volume size can be used to single out dynamics of the system for not too small volumes.

In contrast to quark models and related approaches, the latest lattice QCD studies allow a direct comparison with observables. For example, in Refs.~\cite{Silvi:2021uya,Bulava:2022vpq} not only the presence of the $\Delta(1232)3/2^+$ was confirmed, but phase shifts were extracted allowing direct access to pole position (i.e., resonance width).   In a first step, energy eigenvalues are extracted from lattice correlation functions by solving the generalized eigenvalue problem (GEVP). These energy eigenvalues form a discrete spectrum --not because resonances appear as bound states but because the problem is calculated in a small cubic volume with periodic boundary conditions. Using the L\"uscher method~\cite{Luscher:1985dn, Luscher:1986pf, Luscher:1990ux}, this information can be used to determine phase shifts at the energy eigenvalues $\{E_L\}$.
As discussed in more detail in Ref.~\cite{Doring:2025sgb}, for an energy eigenvalue $E$ ($E=W=\sqrt{s}$ is the total scattering energy) the phase shift $\delta$ at this energy is obtained as
\begin{align}
    q_{\rm cm}(E)\cot\vertarrowbox{\delta}{
            \fbox{\begin{tabular}{c}
                Experiment\\Effective Field Theories\\\ldots
            \end{tabular}
        }}(E)=(-8\pi E)G_L(E)
        ~~\text{for}~~
        E\in\vertarrowbox{\{E_L\}}{\fbox{lattice QCD}}
\label{eq:fv-scattering-5}
\end{align}
with center-of-mass momentum $q_\text{cm}$. It is notable that the right-hand side is proportional to the Lüscher zeta-function ${\cal Z}_{00}$ up to regular and, thus, exponentially suppressed terms $e^{-ML}$,
\begin{align}
    (-8\pi E)G_L(E)
    =
    \frac{2}{\sqrt{\pi}L}\,{\cal Z}_{00}(1,\hat q^2)+\mathcal{O}(e^{-ML})\,,
\label{eq:fv-scattering-6}
\end{align}
where $\hat q=q_{\rm cm}(E)L/(2\pi)$ and $L$ is the side length of the cube.
See Refs.~\cite{Beane:2003yx,Doring:2011vk} for an explicit derivation. 
For a better understanding it is worth considering the infinite-volume scattering problem in which the scattering amplitude can be obtained  
as a solution of
\begin{align}
    T&=V+VGT
     =\frac{1}{V^{-1}-G}
    \quad
    \text{with}
    \quad
    G=\int
    \frac{\dv{^}3\bm{q}}{(2\pi)^3}\frac{1}{2\omega_1(\bm{q})\,\omega_2(\bm{q})}
    \frac{\omega_1(\bm{q})+\omega_2(\bm{q})}
    {E^2-(\omega_1(\bm{q})+\omega_2(\bm{q}))^2+i\epsilon}\,,
    \label{eq:fv-scattering}
\end{align}
for particle masses $m_1,\,m_2$ with corresponding energies $\omega_{1,2}(\bm{q})=\sqrt{m_{1,2}^2+\bm{q}^2}$. 
Note that this corresponds to the LSE of \cref{eq:7.179} with the following modifications: 1) The energy is relativized; 
2) The definition of $T$ has changed from its quantum mechanical normalization to a normalization used in quantum field theory;
3) The only momentum dependence is in $G$, not in $V(E)$ or $T(E)$. This is referred to as ``on-shell factorization''. As  side remarks, this equation respects two-body unitarity~\cite{Mai:2025wjb}; if one neglects Re~$G$, sometimes referred to as the ``dispersive part'', the amplitude is still unitary and referred to as $K$-matrix approach.
Notably, in \cref{eq:fv-scattering} the interaction is parametrized in a real-valued term $V$ while the imaginary part of the amplitude responsible for unitarity arises from $G$.
We note that the loop function $G$ is a relativistic form of the integral over the non-relativistic Green's function $1/(E-H_0)$ and thus $T\sim 1/(E-H)$.

Consider now the same system in a finite volume or, more specifically, a 3-dimensional cube with side length $L$ with periodic boundary conditions. In that system momenta become discretized, i.e., $\bm{q}\in\mathds{R}^3\to \bm{q}\in\aleph_L:=\{(2\pi/L)\bm{n}|\bm{n}\in\mathds{Z}^3\}$ such that the loop function in \cref{eq:fv-scattering} becomes
\begin{align}
    G\to \tilde G=\frac{1}{L^3}\sum_{\bm{q}\in\aleph_L}
    \frac{1}{2\omega_1(\bm{q})\,\omega_2(\bm{q})}\,\,
    \frac{\omega_1(\bm{q})+\omega_2(\bm{q})}
    {E^2-(\omega_1(\bm{q})+\omega_2(\bm{q}))^2}\,.
    \label{eq:fv-scattering-3}
\end{align}
Note that $\tilde G$ is a real but singular function, with poles emerging whenever the on-shell condition $E=\omega_1(\bm{q})+\omega_2(\bm{q})$ is fulfilled for any $\bm{q}\in\aleph_L$.
As expected the difference between $\tilde G$ and $G$ provides the translation from finite to infinite volume, $G_L=\tilde G-\Re G$ with $G_L$ on the right-hand side of the L\"uscher equation~\eqref{eq:fv-scattering-5}.

There are several practical limitations to the approach, all of which can, in principle, be improved upon. Oftentimes unphysical quark and, hence, hadron masses are used such as in Ref.~\cite{Silvi:2021uya} in which the pion is about twice as heavy as the physical one. However, this also represents an advantage for light baryon spectroscopy: The larger mass implies that the elastic window between the $\pi N$ and the $\pi\pi N$ threshold is larger. Three- and more body effects at higher energies are difficult to manage, not only due to conceptual challenges of finite-volume effects, but also to the many combinations in which two pions and a nucleon can be combined to form a given, overall $J^P$. There are not only different $2+1$ combinations available, such as $\pi N^*$, $\pi\Delta$, $(\pi\pi)N$, the latter representing two-pion isobars with spectator nucleon. In addition, both isobars and spectator can have spin. For example, there are up to two $\pi\Delta$ and three $\rho N$ channels for a given $J^P$, see \cref{tab:couplscheme}. 
\begin{table}[htb]
\caption{Angular momentum structure of  coupled channels in isospin $I=1/2$ up to $J=9/2$ as included in the JBW and ANL-Osaka approaches. The $I=3/2$ sector is similar up to obvious isospin selection rules.
In the table, the total spin $S=|\bm{ S}_N+\bm{ S}_\rho|$ is given by the sum of the $\rho$ spin and the nucleon spin, and $L$ is the orbital angular momentum. The $\omega N$ channel has the same coupling scheme as the shown $\rho N$ channel.
Table from Ref.~\cite{Doring:2025sgb}.}
\begin{center}
\begin{tabularx}{0.99\linewidth}{lX|ll|ll|ll|ll|ll}
    \hline \hline
    $\mu$	&\multicolumn{1}{r}{$J^P=$}
    &$\frac{1}{2}^-$&$\frac{1}{2}^+$&$\frac{3}{2}^+$&$\frac{3}{2}^-$&$\frac{5}{2}^-$&$\frac{5}{2}^+$&$\frac{7}{2}^+$&$\frac{7}{2}^-$&$\frac{9}{2}^-$&$\frac{9}{2}^+$\BBB \TT
    \\
    \hline
    $1$	&$\pi N$ 			&$S_{11}$		&$P_{11}$		&$P_{13}$		&$D_{13}$		&$D_{15}$		&$F_{15}$		&$F_{17}$		&$G_{17}$		&$G_{19}$		&$H_{19}$		\bigstrut[t]\\
    $2$	&$\rho N(S=1/2)$		&$S_{11}$		&$P_{11}$		&$P_{13}$		&$D_{13}$		&$D_{15}$		&$F_{15}$		&$F_{17}$		&$G_{17}$		&$G_{19}$		&$H_{19}$		\\
    $3$	&$\rho N(S=3/2, |J-L|=1/2)$	&---		&$P_{11}$		&$P_{13}$		&$D_{13}$		&$D_{15}$		&$F_{15}$		&$F_{17}$
    &$G_{17}$		&$G_{19}$		&$H_{19}$		\\
    $4$	&$\rho N(S=3/2, |J-L|=3/2)$	&$D_{11}$		&---		&$F_{13}$		&$S_{13}$		&$G_{15}$		&$P_{15}$		&$H_{17}$
    &$D_{17}$		&$I_{19}$		&$F_{19}$		\\
    $5$	&$\eta N$ 			&$S_{11}$		&$P_{11}$		&$P_{13}$		&$D_{13}$		&$D_{15}$		&$F_{15}$		&$F_{17}$		&$G_{17}$		&$G_{19}$		&$H_{19}$		\\
    $6$	&$\pi\Delta (|J-L|=1/2)$	&---		&$P_{11}$		&$P_{13}$		&$D_{13}$		&$D_{15}$		&$F_{15}$		&$F_{17}$
    &$G_{17}$		&$G_{19}$		&$H_{19}$		\\
    $7$	&$\pi\Delta (|J-L|=3/2)$	&$D_{11}$		&---		&$F_{13}$		&$S_{13}$		&$G_{15}$		&$P_{15}$		&$H_{17}$
    &$D_{17}$		&$I_{19}$		&$F_{19}$		\\
    $8$	&$\sigma N$			&$P_{11}$		&$S_{11}$		&$D_{13}$		&$P_{13}$		&$F_{15}$		&$D_{15}$		&$G_{17}$		&$F_{17}$		&$H_{19}$		&$G_{19}$		\\
    $9$	&$K\Lambda$ 			&$S_{11}$		&$P_{11}$		&$P_{13}$		&$D_{13}$		&$D_{15}$		&$F_{15}$		&$F_{17}$		&$G_{17}$		&$G_{19}$		&$H_{19}$		\\
    $10$	&$K\Sigma$ 			&$S_{11}$		&$P_{11}$		&$P_{13}$		&$D_{13}$		&$D_{15}$		&$F_{15}$		&$F_{17}$		&$G_{17}$		&$G_{19}$		&$H_{19}$
    \BBB
    \\
    \hline \hline
\end{tabularx}
\end{center}
\label{tab:couplscheme}
\end{table}

Another challenge lies in the sparsity of the finite-volume spectrum. While larger boxes can host more energy eigenvalues, their simultaneous extraction by solving the GEVP becomes more challenging. 
Fewer eigenvalues at higher than physical pion masses is a particular difficult situation, because then resonances are narrower and, e.g., the variation of the $\Delta(1232)3/2^+$~\cite{Silvi:2021uya} phase from $0^\circ$ to $180^\circ$ occurs over a very narrow energy interval. 
A remedy is to consider moving frames~\cite{Rummukainen:1995vs}. In this situation  energy eigenvalues are determined for the scattering system moving relative to the lattice rest frame, resulting in additional energies that can help populate energy gaps. The downside of this method is that different partial waves contribute to the same eigenvalue. The reason lies in the breakdown of rotational symmetry in the cubic volume. While this is a problem, in general, it becomes more severe for moving frames. 

An additional source of partial wave mixing for excited baryons comes from different parities for a given orbital angular momentum, $J=L\pm 1/2$, allowing for more than one $P$-wave, $D$-wave etc. even in infinite volume. For the scattering of spinless mesons this is less of an issue. Finally, the determination of the energy eigenvalues themselves is more difficult for baryons than for mesons due to the increased signal-to-noise ratio in baryon correlation functions. This problem  exists for the ground state baryon and becomes more severe for the higher excited states relevant for light-baryon spectroscopy. As the signal-to-noise ratio diminishes over Euclidian time it becomes increasingly difficult to isolate excited states reliably from even higher, faster decaying excited states. This introduces systematic uncertainties. Also, the calculation of correlation functions is more complicated for three than for two quarks. All these effects combined explain while, despite the recent successes in lattice spectroscopy of two- and three-meson systems, light baryon spectroscopy from the lattice is reaching quantitative levels only recently and remains one of the exciting current developments in nuclear physics. 

To summarize developments in the light-baryon sector,
 see Refs.~\cite{Khan:2020ahz, Engel:2013ig, Dudek:2012ag, Edwards:2011jj, Bulava:2010yg, Melnitchouk:2002eg}.
 More recent approaches have enabled the extraction of observables such as scattering lengths or the $\Delta$ resonance phase shifts mentioned at the beginning of this section. Among those are Refs.~\cite{Pittler:2025upn, BaryonScatteringBaSc:2023ori, BaryonScatteringBaSc:2023zvt, Bulava:2022vpq, Silvi:2021uya, Pittler:2021bqw, Lang:2016hnn, Lang:2012db, Lin:2008qv}, as well as pioneering studies on higher excited baryons~\cite{Owa:2025mep, Zhuge:2024iuw, Liu:2023xvy, Hockley:2023yzn, Abell:2023nex, Wu:2017qve, Wu:2016ixr, Liu:2016wxq, Liu:2016uzk, Liu:2015ktc, Kiratidis:2015vpa}. We refer to dedicated reviews and whitepapers for more details~\cite{Bulava:2022ovd, Padmanath:2018zqw, Briceno:2017max}.
For the extraction of Low Energy Constants from global fits to baryon lattice QCD data, see Refs.~\cite{Lutz:2024ubv, Hudspith:2024kzk} as well as the dedicated sections in the FLAG report~\cite{FlavourLatticeAveragingGroup:2019iem}.


\subsection{Chiral unitary approaches}
(Excited) Meson and (excited) baryon degrees of freedom form the basis for  Chiral Perturbation Theory (CHPT) approaches~\cite{Scherer:2009bt} to excited baryons. See, for example, the reviews by Bernard, Kaiser, and Mei{\ss}ner \cite{Bernard:1995dp},  by Pascalutsa, Vanderhaeghen, and Yang~\cite{Pascalutsa:2006up} on the $\Delta(1232)$ resonance, and by Bernard ~\cite{Bernard:2007zu}. 

Through the unitarization of a chiral (or other) interactions to a given order~\cite{Guo:2017jvc}, light baryons or light-strange baryons can emerge. There is an enormous body of literature on the subject, see Ref.~\cite{Doring:2025sgb} for a comprehensive reference collection. 
Here, we only mention that CHPT provides interactions across mesons and baryons of different strangeness. Indeed, it can be used to provide a unified and consistent picture for the nonstrange meson-baryon S-wave sector the enigmatic  $\Lambda(1405)$ states~\cite{Lu:2022hwm}. 

In a nutshell, in their simplest form, chiral unitary approaches consider the scattering equation~\eqref{eq:fv-scattering} that relates an interaction $V$ derived from CHPT to the scattering matrix $T$ in a nonlinear way, with the formal solution $T=(1-VG)^{-1}V$. The denominator $1-V(E)G(E)$ may become singular for complex scattering energies $E$, leading to a resonance emerging from the non-linear re-summation of the interaction.

The concept of coupled channels can be easily incorporated into the approach, rendering the quantities $V$, $G$, and $T$ matrices in channel space with $V$ providing channel transitions and the diagonal matrix $G$ providing the meson-baryon propagation. As an upside, this procedure provides not only the emergence of some resonances, but also the incorporation of dynamical, non-resonant effects such as threshold cusps. 

As an example, the formalism of Refs.~\cite{Borasoy:2007ku, Bruns:2010sv} relies on the fully covariant Bethe-Salpeter equation (BSE)~\cite{Salpeter:1951sz}. We denote the in- and outgoing meson momenta by $q_1$ and $q_2$, and the overall four-momentum by $p=q_1+p_1=q_2+p_2$, where $p_1$ and $p_2$ are the momenta of in- and out-going baryon, respectively. For the meson-baryon scattering amplitude $T(\slashed{q}_2, \slashed{q}_1; p)$ and chiral kernel $V(\slashed{q}_2, \slashed{q}_1; p)$ the integral equation to solve reads 
\begin{align}
\label{eqn:BSE}
	T(\slashed{q}_2, \slashed{q}_1; p)= V(\slashed{q}_2, \slashed{q}_1; p) 
	+
	i\int\frac{d^d l}{(2\pi)^d}V(\slashed{q}_2, \slashed{l}; p) 
	S(\slashed{p}-\slashed{l})\Delta(l)T(\slashed{l}, \slashed{q}_1; p) \ ,
\end{align}
where $S$ and $\Delta$ represent the baryon (of mass $m$) and the meson (of mass $M$) propagator respectively, that are given by $iS(\slashed{p}) ={i}/({\slashed{p}-m+i\epsilon})$ and $i\Delta(k) ={i}/({k^2-M^2+i\epsilon})$. The suppressed channel indices in the above formulas include all relevant (meson-baryon) coupled channels, such that $T$, $V$, $S$ and $\Delta$ are matrices in channel space. For strangeness $S=0$ and electric charge $Q=+1\,$ one is left with the following channels: $\{p \pi^0,~n \pi^{+},~p\eta,~\Lambda K^+,~\Sigma^0 K^+,~\Sigma^+ K^0\}$. Separating the momentum space from channel space structures the chiral potential considered here takes the form:
\begin{align}
\label{eqn:coupling}
	V(\slashed{q}_2, \slashed{q}_1; p)=
    \underbrace{
        A_{WT}(\slashed{q_1}+\slashed{q_2})}_{\text{leading order CHPT}}
	+\underbrace{
        A_{14}(q_1\cdot q_2)+A_{57}[\slashed{q_1},\slashed{q_2}]+A_{M}(q_1\cdot q_2)+A_{811}\big(\slashed{q_2}(q_1\cdot p)+\slashed{q_1}(q_2\cdot p)\big)}_{\text{next-to-leading order CHPT (local terms)}}
    \,,
\end{align}
where the first matrix ($A_{WT}$) only depends on the meson decay constants $F_\pi ,\,F_K,\,F_{\eta}$, whereas  $A_{14}$, $A_{57}$, $A_{811}$ and $A_{M}$ also depend on the NLO low-energy constants as specified in the appendix of Refs.~\cite{Mai:2012wy,Mai:2013cka}. The loop diagrams appearing in the BSE Eq.~(\ref{eqn:BSE}) are in general divergent and require renormalization. In case of a strict chiral perturbation expansion, the terms can be renormalized  in a quite straightforward way, order by order, including, at a given order, all the counterterms absorbing the loop divergencies. With this formalism, in Ref.~\cite{Bruns:2010sv} the low energy constants and and regularization scales are fitted to the $S_{11}$ and $S_{31}$ partial wave, resulting in the emergence of the $N(1535) 1/2^-$ resonance and a second peak that could be interpreted as the prediction of the $N(1650)1/2^-$, confirming and going beyond earlier work on theses states and their chiral dynamics~\cite{Kaiser:1995cy, Inoue:2001ip, Nieves:2001wt, Lutz:2005ip}.

As the CHPT interactions arises from a field theory, the coupling of weak probes to the hadronic components is known. This property can be used for extensions of the approach to  manifestly gauge-invariant photoproduction amplitudes~\cite{Borasoy:2007ku, Doring:2007rz, Mai:2012wy,Mai:2013cka}.
The description of some light baryons as dynamically generated states can be tested by comparing their predicted electromagnetic properties~\cite{Bruns:2022sio,  Mai:2012wy, Ruic:2011wf, Doring:2010fw, Doring:2010rd, Doring:2009qr, Roca:2006pu, Doring:2005bx, Kaiser:1996js} to recent data and phenomenological analyses discussed in Sec.~\ref{sec:em}.
Analogously, coupling vector mesons to constituents allows for predictions of the corresponding reaction cross sections~\cite{Khemchandani:2012ur, Geng:2008cv, Doring:2008sv}.
Rescattering through chiral unitary potentials can be extended to three particles~\cite{Malabarba:2021taj, Khemchandani:2020exc, MartinezTorres:2010zv,  MartinezTorres:2007sr}. There are reviews on coupled-channel scattering by Hyodo~\cite{Hyodo:2011ur}, Mai~\cite{Mai:2020ltx}, Mei{\ss}ner~\cite{Meissner:2020khl}, Oller~\cite{Oller:2000ma}, and Oset~\cite{Oset:2016lyh} with emphasis on chiral unitary methods and, at least partially, covering light baryons. 


\section{Phenomenology of light baryon resonances}

\subsection{Baryon amplitudes for resonance extraction}

As mentioned before, light baryon resonances are fully characterized by spin-parity $J^P$ and isospin, $N$ ($\Delta$) for $I=1/2$ ($I=3/2$). For most values of $J$ there are two possibilities for the orbital angular momentum $L$ of the $\pi N$ system, $J=L\pm1/2$ ($J\ge1/2$) that is the historically most studied channel for light baryons.
When kinematically possible, resonances also appear in the production of heavier two-body systems, such as $\eta N$, $K\Lambda$, $K\Sigma$,\dots.
Both mesonic and baryonic part of the decay products can have a more complicated spin structure. For example,
there are different possibilities for  $\pi\Delta$ and $\rho N$ channels to couple to a given $J^P$; each of those possibilities is counted  as individual channels. As an example, we show the  coupling scheme of the J\"ulich-Bonn-Washington (JBW) approach~\cite{Ronchen:2012eg} in Table~\ref{tab:couplscheme}. The ANL-Osaka (former EBAC) approach has a very similar channel space~\cite{Kamano:2013iva, Kamano:2019gtm, Kamano:2010ud, Kamano:2009im, Suzuki:2009nj, Julia-Diaz:2009dnz, Kamano:2008gr}, see Ref.~\cite{Matsuyama:2006rp} for details on the formalism.

The unstable ``isobars'' in the $\pi\sigma$, $\pi\Delta$, and $\rho N$ channels stand for two-body sub-channels with sub-energy-dependent amplitude rather than simple resonances. The extension of the original JBW model to include the $K\Lambda$ and $K\Sigma$ channels was carried out in Ref.~\cite{Doring:2010ap}, while the $\omega N$ channel was added only recently by Wang et al.~\cite{Wang:2022osj}.
The $\omega N$ channel was added to the ANL-Osaka model in Ref.~\cite{Paris:2008ig}.

At the heart of these approaches is a scattering equation~\cite{Krehl:1999km, Gasparyan:2003fp, Doring:2009bi, Doring:2009yv, Doring:2010ap, Ronchen:2012eg, Ronchen:2014cna, Ronchen:2015vfa, Ronchen:2018ury}  that shows substantial similarities with the aforementioned scattering equations \eqref{eq:7.179} and \cref{eq:fv-scattering},
\begin{align}
    T_{\mu\nu}(p',p;E)=V_{\mu\nu}(p',p;E)
    +\sum_{\kappa}\int\limits_0^\infty \dv{q}\,
     q^2\,V_{\mu\kappa}(p',q;E)\,G_\kappa(q;E)\,T_{\kappa\nu}(q,p;E) \ , 
    \label{scattering}
\end{align}
where $p'\equiv|\bm{p}\,'|$ ($p\equiv |\bm{p}\,|$) is the modulus of the outgoing (incoming)  three-momentum that may be on- or off-shell, $E=\sqrt{s}$ is the scattering energy, and $\mu,\,\nu,\,\kappa$ are channel indices. In
Eq.~(\ref{scattering}), the propagator for a stable meson and a stable baryon is given by 
\begin{align}
    G_\kappa=\frac{1}{E-E_\kappa(q)-\omega_\kappa(q)+i\epsilon}\ ,
    \label{gkappa}
\end{align}
with meson and baryon energies $\omega_\kappa$ and $E_\kappa$, respectively. 
The propagators for three-body channels are more complicated~\cite{Doring:2025sgb}.

\begin{figure}[htb]
    \centering
    \includegraphics[width=0.6\textwidth,angle=-0]{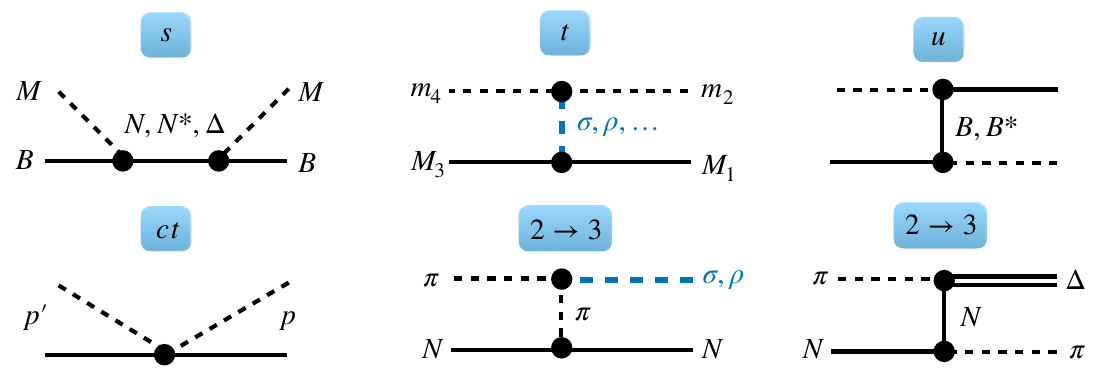}
    \caption{
    Meson ($M$) baryon ($B$) transitions through $s$-, $t$-, $u$-channel, and contact $(ct)$ processes. Also, the $2\to 3$ processes to populate the effective  $\sigma N$, $\rho N$, and $\pi\Delta$ channels in the ANL-Osaka and the JBW approaches are shown. The figure also labels incoming (outgoing ) c.m. momenta $p$ ($p'$) and baryon (meson) masses $M_1,\,M_3$ ($m_2,\,m_4$). Figure from Ref.~\cite{Doring:2025sgb}.}
    \label{fig:alldiagrams}
\end{figure}

\begin{figure}[htb]
    \centering
     \includegraphics[width=0.41\linewidth]{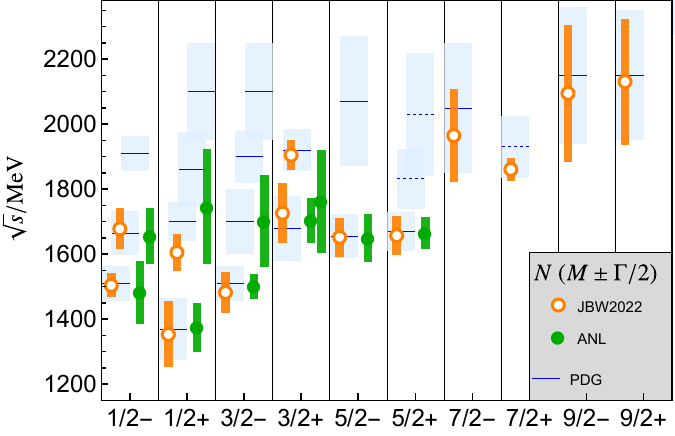}
    ~~~~
    \includegraphics[width=0.41\linewidth]{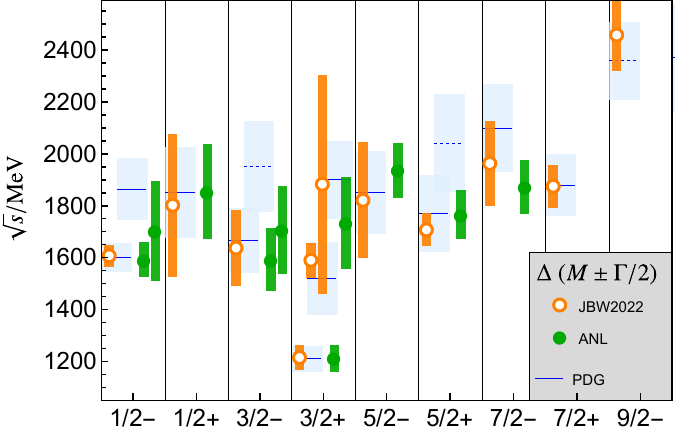}
    \caption{Summary of results for the baryon parameters obtained in the ANL-Osaka~\cite{Kamano:2013iva} and JBW~\cite{Ronchen:2022hqk} approaches. The lengths of the bars correspond to the resonance widths. For these cases, PDG information~\cite{ParticleDataGroup:2024cfk} is displayed in blue. Note that for the ANL-Osaka model, only resonances up to $2\,\GeV$ in mass and $400\,\MeV$ in width are quoted.
    Pictures from Ref.~\cite{Doring:2025sgb}.}
    \label{fig:BaryonsSummary}
\end{figure}

The  $t$- and $u$-channel exchanges included by the ANL-Osaka and JBW approaches are shown in Fig.~\ref{fig:alldiagrams}
\begin{table}[htb]
\caption{ The $t$- and $u$-channel transitions in the JBW and ANL Osaka models. Both models also include $s$-channel poles and contact interactions that are not listed here.
Black entries show common transitions, blue (green) transitions indicate processes only contained in the JBW model but not the ANL-Osaka model (ANL-Osaka model but not JBW model). The particle type implies which interaction channel ($t$ or $u$) is meant. Abbreviations used: $\Delta$ for $\Delta(1232)3/2^+$, $\sigma$ for $f_0(500)$, $\kappa$ for $K_0^*(700)$, $\rho$ for $\rho(770)$, $a_0$ for $a_0(980)$, $f_0$ for $f_0(980)$, $\Sigma^*$ for $\Sigma(1385)3/2^+$, $\Xi^*$ for $\Xi(1530)3/2^+$, and $C$ for contact term. The $(\pi\pi)_\sigma$ and $(\pi\pi)_\rho$ entries correspond to correlated two-pion exchange that is treated differently in both approaches. 
Table from Ref.~\cite{Doring:2025sgb}.}
\begin{center}
\begin{tabularx}{0.99\linewidth}{l|XXXXXXXX} 
\hline \hline
$\mu$ & $\pi N$ & $\eta N$ & $K\Lambda$ & $K\Sigma$ & ${\color{blue}\omega N}$ & $\pi \Delta$ & $\sigma N$ & $\rho N$\TT\\
\hline
\rowcolor{lightgray}
    $\pi N$ & \makecell{$N$, $\Delta$,\\
    {\color{blue}$(\pi\pi)_\sigma$}, {\color{teal} $\sigma, f_0$}\\ 
    {\color{blue}$(\pi\pi)_\rho$}, {\color{teal} $\rho$}}
    & {\color{blue} $a_0$}, $N$                        
    & \makecell{$K^*$, $\Sigma$,\\ $\Sigma^*$
    ,\\ {\color{teal} $\kappa$} }\ \ \ 
    & \makecell{$K^*$,  $\Lambda$,\\ $\Sigma$, $\Sigma^*$
    ,\\ {\color{teal} $\kappa$} } \  
    & {\color{blue} $\rho$, $N$}
    & $\rho$, $N$, $\Delta$ 
    & $\pi$, $N$ 
    & \makecell{$\pi$, $\omega$,\\ {\color{blue}$a_1$}, $N$
    ,\\ {\color{blue} $\Delta$}, $C$}
\\
$\eta N$ &&
    $N$, {\color{blue}$f_0$}
    & \makecell{$K^*$, $\Lambda$,\\ {\color{teal} $\kappa$}} 
    & \makecell{$K^*$, $\Sigma$
    ,\\ {\color{blue}$\Sigma^*$}, {\color{teal}$\kappa$}} 
    & {\color{blue} $\omega$, $N$} 
    &   
    & {\color{teal}$N$}  
    & {\color{teal}$N$}  
\\ \rowcolor{lightgray}
$K\Lambda$ &&&
     \makecell{$\omega$, {\color{blue}$f_0$}, $\phi$,\\ $\Xi$, {\color{blue}$\Xi^*$} } 
    & \makecell{$\rho$, {\color{blue}$a_0$},\\ $\Xi$, {\color{blue}$\Xi^*$}}\ \ \ \ 
    & \makecell{ {\color{blue}$K$, $K^*$},\\ {\color{blue}$\Lambda$} } 
&&& \\
$K\Sigma$ &&&&  
     \makecell{$\rho$, $\omega$, $\phi$,\\ {\color{blue} $f_0$, $a_0$},\\ $\Xi$, {\color{blue}$\Xi^*$}} 
     & \makecell{ {\color{blue}$K$, $K^*$},\\{\color{blue} $\Sigma$, $\Sigma^*$} } 
\\ \rowcolor{lightgray}
{\color{blue}$\omega N$} &&&&&
    {\color{blue}$\sigma$, $N$} &&&
\\
$\pi\Delta$ &&&&&&
     $\rho$, {\color{blue} $N$}, $\Delta$ 
     & {\color{blue}$\pi$} 
     & $\pi$, $N$
\\ \rowcolor{lightgray}
$\sigma N$ &&&&&&&
    {\color{blue}$\sigma$}, $N$ &{\color{teal} $N$} 
\\
$\rho N$ &&&&&&&& \makecell{{\color{blue}$\rho$}, $N$,\\ $\Delta$, $C$}
\BBB
\\
\hline \hline
\end{tabularx}
\end{center}
\label{tab:trans}
\end{table}
  and compared in Table~\ref{tab:trans}, which also contains the list of included channels.
In Fig.~\ref{fig:BaryonsSummary} the baryon spectrum is shown as determined by the ANL-Osaka~\cite{Kamano:2013iva}
and the JBW~\cite{Ronchen:2022hqk} approaches, compared to the spectrum from the PDG.

There are many other approaches dedicated to the analysis of pion- and photon-induced reactions, ranging from fully covariant approaches as in \cref{eqn:BSE}, approaches that are made covariant as in Ref.~\cite{Huang:2011as}, to $K$-matrix approaches including dispersive parts, as the Bonn-Gatchina (BnGa) approach~\cite{Anisovich:2011fc} or the Chew-Mandelstam formulation used in the SAID analysis~\cite{Workman:2012jf}. Notably, the BnGa approach includes the largest number of channels and data in their analysis. All mentioned approaches help improve the determination of light baryon resonance parameters as reflected in the PDG. For a more detailed discussion including a comprehensive literature review, see Ref.~\cite{Doring:2025sgb}.

\subsection{Analytic structure of partial waves}
\begin{figure}[htb]
\begin{center}
 \includegraphics[width=0.5\textwidth]{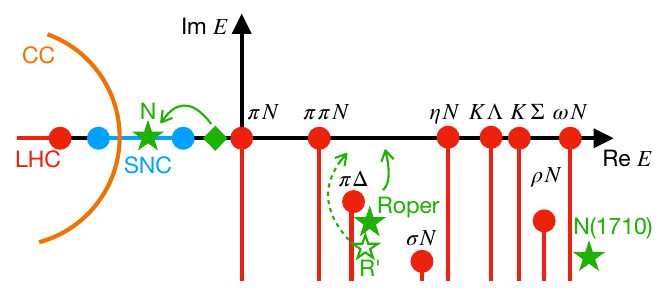}
\end{center}
\caption{Analytic structure of the $P_{11}$ partial wave, for complex scattering energies $E=W$. The circular
cut is labeled ``CC'', the left-hand cut ``LHC''. The nucleon is labeled as $N$.  See text for further explanations. Picture from Ref.~\cite{Doring:2025sgb}.}
\label{fig:anap11}     
\end{figure}

Not every bump in a cross section or a partial wave is a resonance. Structures that are not resonances but arise from scattering kinematics need to be identified and, if possible, included in the formulation of amplitudes to avoid misidentification as resonances. In terms of the scattering amplitude as a complex function of energy and momenta, such terms are non-analyticities other than simple poles. In the light baryon sector, there are, on one hand, 2-body thresholds from states such as $\eta N$, $K\Lambda$, $K\Sigma$, or $\omega N$ that can be included in the amplitude through coupled channels. For an illustration, see the analytic structurxe of the $P_{11}$ partial wave in Fig.~\ref{fig:anap11}.
The thresholds of these channels appear as branch points on the real-energy axis. Each of these two-body channels is associated with one  cut (red lines). 

Then there are three-body channels. As mentioned in Sec.~\ref{sec:lQCD}, they can be organized in isobar-spectator channels in which the isobar symbolized the correlation of two strongly interacting particles. In general, one has the discussed $\pi N^*$, $\pi\Delta$, $N(\pi\pi)$ spectator-isobar combinations (in a restriction to non-strange isobars). Input for the isobars is usually obtained through amplitudes that were fitted to the corresponding two-body processes~\cite{Schutz:1994ue}.
Figure~\ref{fig:anap11} shows the $\pi\pi N$ branch points on the real axis arising from the three-body channels, plus branch points in the complex plane from the three 3-body channels, $\pi\Delta$, $\sigma N$, and $\rho N$. The role of these structure as a possible source of misidentification of resonances is discussed in Ref.~\cite{Ceci:2011ae}. 

To search for poles on the most relevant Riemann sheets, the cuts with thresholds are usually chosen to be oriented from the threshold openings into the negative imaginary $E$-direction as indicated in the figure. In addition, the figure shows the sub-threshold short nucleon cut (SNC), circular cut (CC) and left-hand cut (LHC). The structure of some of these cuts and branch points is discussed in Ref.~\cite{Hohler:1984ux}, see also the Appendix of Ref.~\cite{Doring:2009yv}. 

The (green) stars in Fig.~\ref{fig:anap11} indicate the positions of the physical poles. The nucleon appears as a bound state pole on the first Riemann sheet in the amplitude, right on top of the SNC. The green diamond indicates the bare nucleon pole.
The green arrow indicates that in 
formulations of the scattering amplitude the nucleon pole is introduced as a bare state, $V\sim g^2/(E-E_0)$, which then undergoes rescattering through \cref{scattering}. This modifies both the pole position (i.e., mass) of the nucleon and its coupling to the $\pi N$ channel. As both quantities are known, the ``bare'' parameters $E_0$ and $g$ can be re-adjusted such that the nucleon pole and its residue acquire physical values in the full scattering amplitude~\cite{Ronchen:2012eg}. This process is referred to as ``renormalization''.

As Fig.~\ref{fig:anap11} shows, the $P_{11}$ partial waves contains the Roper resonance $N(1440)1/2^+$ as an excited state of the nucleon with the same quantum numbers. Not only that, but 
there are two Roper poles on different $\pi\Delta$ sheets, labeled as $N(1440)$ and $N^\prime(1440)$ (not to be confused with the two-pole structure of the $\Lambda(1405)$ for which both poles are on the \emph{same} Riemann sheet). This two-pole structure is found in several analyses, such as GWU/SAID~\cite{Arndt:2006bf}, EBAC~\cite{Suzuki:2009nj}, and JBW~\cite{Ronchen:2012eg}. It is simply due to the fact that poles can repeat on different sheets. In addition, the proximity of the complex $\pi\Delta$ branch point to the resonance poles leads to a non-standard shape of the Roper resonance that is not of the Breit-Wigner type. Another reason for the unusual shape is the strong influence of the $\sigma N$ channel in which all three particles $\pi$, $\pi$, $N$ are in relative $S$-waves for $P_{11}$. The absence of any centrifugal barriers leads to a large inelasticity at low energies, strongly distorting the Roper resonance shape. In addition, other resonances  like the $N(1710)1/2^+$  have been reported in most analyses as the figure indicates. 

The list of non-resonant bump structures wouldn't be complete without mentioning triangle singularities. See Refs.~\cite{Bayar:2016ftu, Mikhasenko:2015oxp, Guo:2019twa} for pedagogical explanations.
The main idea is that in an isobar channel with a very narrow resonance and a stable spectator ``1'' the two states can be onshell at a certain momentum $p$ that depends on overall energy $E$. The narrow resonance decays into two particles 2 and 3, and 2 recombines with the initial spectator 1 to another isobar. In backward kinematics for the exchanged particle 2, sometimes there are mass combinations for which particles 1, 2, 3, and the narrow resonance can be onshell for a certain $E$. This results in a kinematic enhancement at that energy referred to as triangle singularity.  For an example in the nonstrange baryon sector see Ref.~\cite{Wang:2016dtb}.
While triangle singularities are usually calculated in a loop diagram, they are, in fact, non-diagonal channel transitions in an infinite series of isobar-spectator rescattering that can be organized in a LSE such as \cref{scattering}. The influence of rescattering effects in such a scheme was studied in Ref.~\cite{Sakthivasan:2024uwd}. 


\subsection{Electromagnetic properties of light baryon resonances}
\label{sec:em}
In the last three decades, a plethora of final states, differential cross sections, and spin observables were measured in photoproduction reactions at JLAB/CLAS~\cite{CLAS:2006pde,  CLAS:2015pjm, CLAS:2015ykk,  CLAS:2024bzi}, CB-ELSA/TAPS~\cite{CBELSATAPS:2007oqn,  CBELSATAPS:2009ntt,  CBELSATAPS:2014wvh,  CBELSATAPS:2021osa} (continued at the new BGOOD detector~\cite{BGO-OD:2019utx, Rosanowski:2024rww,  Jude:2022atd, BGOOD:2021sog, Alef:2020yul}),
MAMI/A2/CrystalBall~\cite{A2:2017gwp, A2:2014pie,  CrystalBallatMAMI:2009lze}, LEPS~\cite{LEPS:2017jqw,  LEPS:2009pib}, 
ELPH/FOREST~\cite{Ishikawa:2019rvz}, and other facilities. Sometimes observables are tied to specific choices of reference frames that are not unique across different experiments and analyses. The ``Rosetta Stone Paper'' provides an overview~\cite{Sandorfi:2011nv}. 
In Ref.~\cite{Anisovich:2016vzt} it is demonstrated how new polarization data leads to converging results among different partial-wave analysis groups.
The unprecedented abundance and precision of data led to the discovery and confirmation of many new states as recently reviewed in Ref.~\cite{Burkert:2025coj}. 

Apart from the above-mentioned SAID, BnGa, JBW, and ANL-Osaka frameworks, one has to mention the MAID approach that provides a unified analysis of pion photo and electroproduction~\cite{Drechsel:2007if} using suitably unitarized Breit-Wigner forms for the resonances. One of several MAID updates, including $\eta$ and $\eta'$ photoproduction, was published in Ref.~\cite{Tiator:2018heh}. There is also the approach of Refs.~\cite{Aznauryan:2014xea, CLAS:2009ces} and the JM reaction model of Refs.~\cite{CLAS:2012wxw, Mokeev:2015lda, Mokeev:2023zhq} to determine baryon resonance properties in electroproduction reactions. 

\begin{figure}[htb]
\begin{center}
 \includegraphics[width=0.45\textwidth]{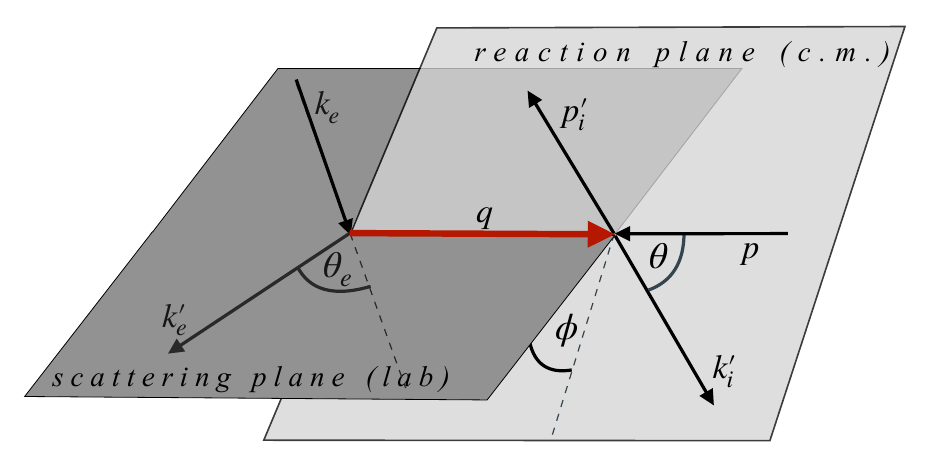}
 \end{center}
 \caption{Kinematics of an electroproduction experiment with
the final meson-baryon state i. The scattering plane is defined
by the respective in/outgoing electron momenta $k_e/k'_e$ with
the electron scattering angle $\theta_e$. The reaction plane is spanned
by the virtual photon and the outgoing meson, scattered by
an angle $\theta$. The momenta $q$ and $p$ correspond to the virtual
photon and target nucleon while $k'_i$ and $p'_i$ correspond to the
outgoing meson and baryon, respectively. Figure from Ref.~\cite{Mai:2023cbp}.}
\label{fig:eeprime}     
\end{figure}

The key electromagnetic properties of light baryon resonances are are their transition form factors (TFFs).
They parametrize the response of the resonance to the excitation with a virtual photon $\gamma^*$, a process referred to as electroproduction~\cite{Carman:2023zke, Mokeev:2022xfo}. The momentum transfer of the photon is $Q^2=-q^2$ with the kinematics illustrated in Fig.~\ref{fig:eeprime}.
Allowing the photon to become virtual enables the study of resonances structure.
For a real photon one has $-Q^2=m_\gamma^2=0$ which allows one to analyze photo- and electroproduction data together. A simultaneous analysis of these reactions, including different final state for electroproduction reactions, was performed in the JBW framework~\cite{Wang:2024byt, Mai:2023cbp, Mai:2021aui, Mai:2021vsw}. 

 Dyson-Schwinger~\cite{Cloet:2008re, Wilson:2011aa, Segovia:2014aza, Segovia:2015hra, Eichmann:2016hgl,  Qin:2019hgk,  Lu:2019bjs}, quark models~\cite{Capstick:1992uc,  Santopinto:2012nq, Golli:2013uha, Aznauryan:2017nkz, Ramalho:2020nwk}, and chiral unitary approaches~\cite{Jido:2007sm, Doring:2010rd} compare or fit their predictions of TFFs to values extracted from data. In this context, remarkable agreement of the lower-lying baryon spectrum with predictions has been achieved~\cite{Eichmann:2016hgl, Qin:2019hgk}, showing little evidence for a ``missing resonance'' problem at lower energies. 
 A key finding concerned the interpretation of the Roper resonance as a radial excitation of the nucleon~\cite{Burkert:2017djo}.
 See Refs.~\cite{Aznauryan:2011qj, Aznauryan:2012ba, Eichmann:2016yit, Eichmann:2022zxn} for reviews and Ref.~\cite{Doring:2025sgb} for a comprehensive literature list.

One can either perform a $Q^2$-independent fit to electroproduction data in fixed $Q^2$ bins or construct an amplitude with explicit $Q^2$ and energy dependence (in both cases the amplitude still depends on the various angles shown in Fig.~\ref{fig:eeprime}). One can then perform energy-dependent Breit-Wigner fits or calculate the $Q^2$-dependent residues at the poles to either get  real-valued BW TFFs or TFFs at the resonance poles. In case of a clear BW resonance the difference is minor. If one has more complicated resonances like the $N(1535)1/2^-$, using the BW definition one might obtain results that depend on the chosen background or the reaction final state under analysis. The real-valued BW TFFs are preferred by many theory groups because their calculations also produce real-valued results, often because baryons might have been calculated as three-body bound states. The resonance pole definition avoids the dependence on background and reaction studied, but then not all theory approaches can compare to the complex-valued TFFs. 

\begin{figure}[htb]
    \raisebox{-.5\height}{\includegraphics[height=5.5cm]{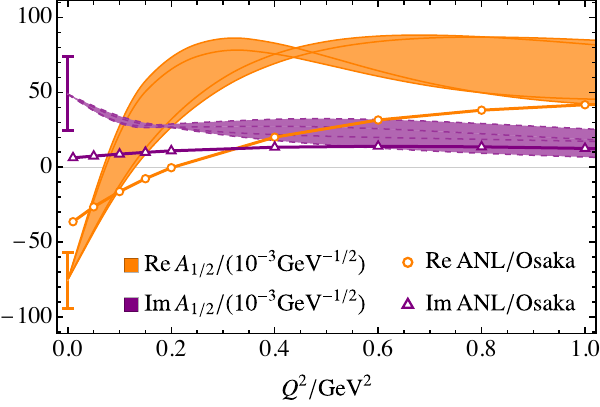}}
    ~~~~~~
    \raisebox{-.5\height}{\includegraphics[height=6.5cm]{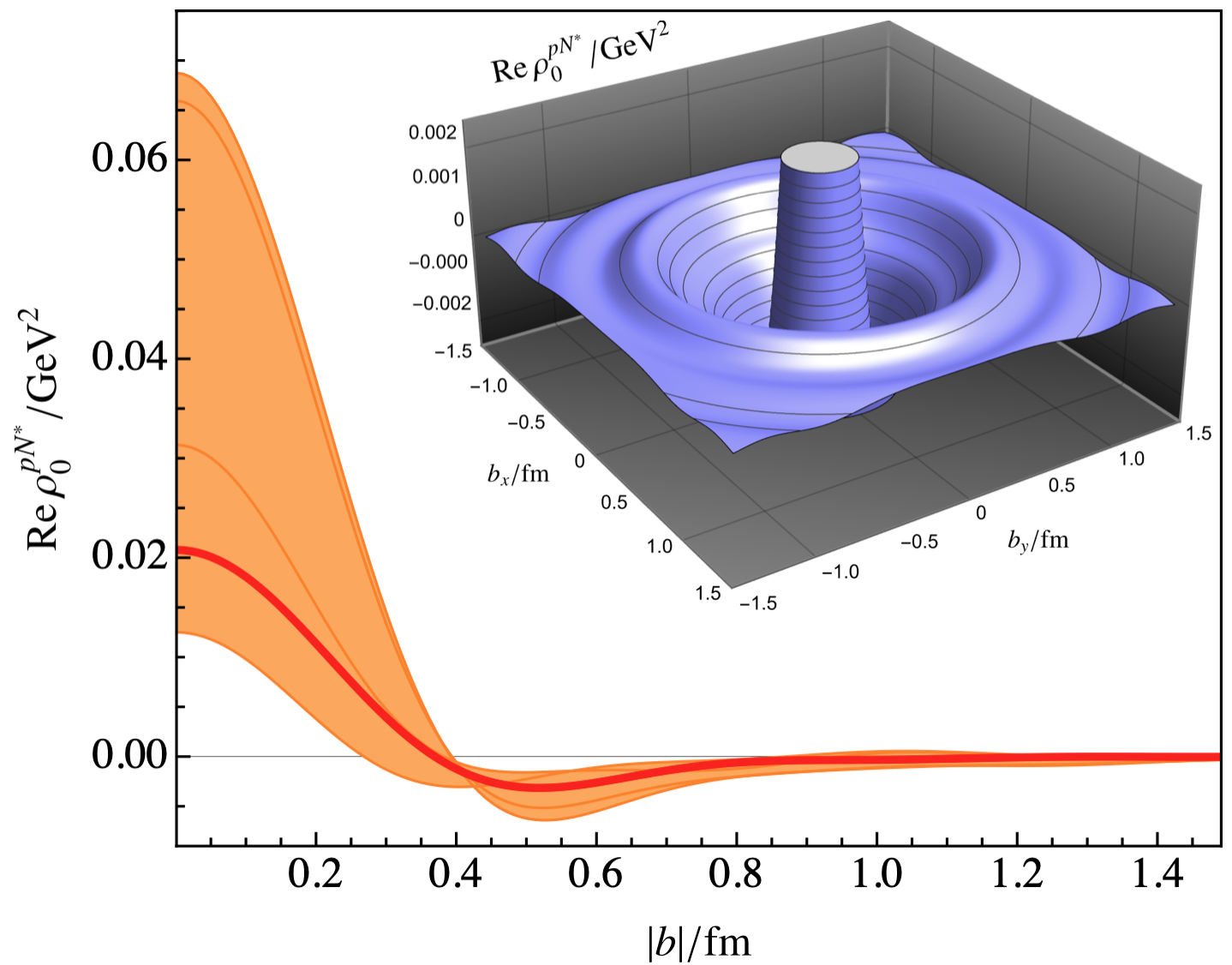}}
	 \caption{
     {\bf Left}: $N(1440)$ transition form factors at small $Q^2$ from the JBW analysis~\cite{Wang:2024byt} in comparison with the ANL-Osaka solutions~\cite{Kamano:2018sfb}. The error bars at $Q^2=0\,\GeV^2$ depict the uncertainties of the photoproduction solution at the pole from Ref.~\cite{Ronchen:2018ury}. A zero in the real part of the TFF is observed.
     {\bf Right}:     
     Unpolarized transverse charge density $\rho_0^{pN^*}$ of the $p\to N(1440)$ transition as a function of the transverse position $b$ in the $xy$-plane from Refs.~\cite{Mai:2023cbp, Wang:2024byt}. The orange band (thick red line) depicts the uncertainty band of the determination (the result using the MAID 2007 helicity couplings~\cite{Drechsel:2007if}). The inset shows the corresponding representation in position space.
     Pictures from Ref.~\cite{Doring:2025sgb}.}
     \label{fig:ChargeDensityRoper} 
\end{figure}

 In the JBW coupled-channel approach, twelve $N^*$ and $\Delta$ transition form factors at the pole were extracted recently in Ref.~\cite{Wang:2024byt} using data with the center-of-mass energy from $\pi N$ threshold to $1.8\,\GeV$, and the photon virtuality $0\leq Q^2/(\GeV/c)^2\leq 8$. Transition form factors at the poles of some higher excited states were estimated for the first time. 
 For the $\Delta(1232)3/2^+$ and $N(1440)1/2^+$ states, the results are in qualitative but not quantitative agreement with previous studies as the left-hand side of Fig.~\ref{fig:ChargeDensityRoper} shows. In particular, one recognizes the unusual zero in the real part of $A^{1/2}$ (albeit at different $Q^2$) that led to the interpretation of the Roper resonance as a radially excited state of the nucleon mentioned before. 
Since quarks are charged, the electroproduction probe can provide a spatial scan of the charge density $\rho$ of, e.g., $p\to N^*$ transition in a certain reference frame~\cite{Tiator:2009mt,Tiator:2008kd,Ramalho:2023hqd}.
This is shown in \cref{fig:ChargeDensityRoper} to the right.

\section*{Acknowledgment}
The work of MD is supported by the National Science Foundation under Grant No. PHY-2310036 and by the U.S. Department of Energy grant DE-SC0016582 and DOE Office of Science, Office of Nuclear Physics under contract DE-AC05-06OR23177. This work contributes to the aims of the U.S. Department
of Energy ExoHad Topical Collaboration, contract DE-SC0023598.
\bibliographystyle{elsarticle-num} 
\bibliography{bibs/biblio,bibs/jh-bb,bibs/jh-mb,bibs/ppnp-ts,bibs/NON-INSPIRE}
\end{document}